\documentclass[12pt]{article}
\usepackage{bbm}
\begin{document}

\title{\normalsize \hfill UWThPh-2002-33 \\[1cm] \LARGE
Maximal atmospheric neutrino mixing\\ in an $SU(5)$ model
}

\author{W.\ Grimus \\
\small Universit\"at Wien, Institut f\"ur Theoretische Physik \\
\small Boltzmanngasse 5, A--1090 Wien, Austria
\\[3mm]
L.\ Lavoura \\
\small Universidade T\'ecnica de Lisboa \\
\small Centro de F\'\i sica das Interac\c c\~oes Fundamentais \\
\small Instituto Superior T\'ecnico, P--1049-001 Lisboa, Portugal}

\date{November 21, 2002}

\maketitle

\begin{abstract}
We show that maximal atmospheric and large solar neutrino mixing can be
implemented in $SU(5)$ gauge theories,
by making use of the $U(1)_F$ symmetry
associated with a suitably defined family number $F$,
together with a $\mathbbm{Z}_2$ symmetry which does not commute with $F$.
$U(1)_F$ is
softly broken by the mass terms of the right-handed neutrino
singlets, which are responsible for the seesaw mechanism;
in addition,
$U(1)_F$ is also spontaneously broken at the electroweak scale.
In our scenario, lepton mixing stems exclusively
from the right-handed-neutrino Majorana mass matrix,
whereas the CKM matrix originates solely in the up-type-quark
sector. We show that, despite the non-supersymmetric character of our model,
unification of the gauge couplings can be achieved at a scale 
$10^{16}\; \mathrm{GeV} < m_U < 10^{19}\; \mathrm{GeV}$;
indeed,
we have found
a particular solution to this problem which
yields results almost identical to the ones of
the Minimal Supersymmetric Standard Model.
\end{abstract}

\newpage

\section{Introduction}

The solar and atmospheric neutrino deficits---for a recent review see,
e.g., Ref.~\cite{concha}---are most naturally explained by neutrino
oscillations \cite{pontecorvo}, with  matter effects
playing a decisive role for solar neutrinos \cite{MSW}. Whereas the
favoured solution of the solar neutrino problem, the large-mixing-angle
MSW solution, displays a large but non-maximal mixing angle
$\theta$,
the atmospheric neutrino problem with mixing angle
$\theta_\mathrm{atm}$ requires
$\sin^2 2\theta_\mathrm{atm} > 0.92$ at 90\% CL \cite{SK}.
It is not difficult to explain large (not necessarily maximal)
atmospheric neutrino mixing---for reviews of mass-matrix textures
for neutrino masses and lepton mixing see Ref.~\cite{reviews}.
However, if the experimental lower bound 
on $\sin^2 2\theta_\mathrm{atm}$ moves closer to 1,
then the need for a symmetry to explain in a natural way
$\theta_\mathrm{atm} \simeq 45^\circ$ becomes acute. 
It has been argued---see for instance
Refs.~\cite{nussinov,wetterich,king}---that such a symmetry should be
non-abelian.
A few papers have attempted to explain nearly maximal atmospheric
neutrino mixing in this way---for an incomplete list of references see 
Refs.~\cite{mohapatra,ohlsson,ma,kitabayashi}. Other approaches to
this problem have also been suggested---for an interesting model 
with lopsided mass matrices see Ref.~\cite{babu}. The
renormalization-group evolution of the lepton mixing angles from the
grand unification scale $m_U$ down to the electroweak scale $m_Z$
(the mass of the $Z$ boson)
has been considered in many papers---see for instance
Refs.~\cite{tanimoto,ellis,lindner} and the works cited
therein. Interesting results can be obtained in this way
\cite{antusch}, with part of the lepton-mixing problem tackled
by the renormalization-group evolution while the residual problem is
left to be solved at the grand unification scale.

In the present paper we discuss the model for maximal
atmospheric neutrino mixing introduced in Ref.~\cite{GL01}, which
belongs to the category of those using a non-abelian symmetry
group (see Section 3 of Ref.~\cite{GL01}).
For simplicity, let us call that model the Maximal
Atmospheric Mixing Model (MAMM). Our aim in this paper is to 
show that the MAMM, which is a simple extension of the Standard Model
(SM), can be embedded in a Grand Unified Theory (GUT)
based on the gauge group $SU(5)$ \cite{su5} (for a textbook see, e.g.,
Ref.~\cite{ross}; for recent papers on the minimal supersymmetrized
$SU(5)$ GUT see Ref.~\cite{bajc}). 
This ``prototype GUT'' can be considered as a
testing ground for ideas on neutrino masses and mixing---for a recent
paper see, e.g., Ref.~\cite{altarelli}.

First we summarize the MAMM. It concerns only the lepton sector of the SM,
with its gauge group which we abbreviate as
\begin{equation}\label{GSM}
G_\mathrm{SM} = SU(3) \times SU(2) \times U(1)\, .
\end{equation}
There are the three well-known lepton families and, in addition, 
three right-handed neutrino singlets $\nu_R$ with a Majorana mass term
\begin{equation} \label{LM}
\mathcal{L}_M = \frac{1}{2}\, \nu_R^T C^{-1} \! M_R^* \nu_R -
\frac{1}{2}\, \bar \nu_R M_R C \bar \nu_R^T\, ,
\end{equation}
where $C$ is the charge-conjugation matrix
and $M_R$ is symmetric. We implement the seesaw mechanism
\cite{seesaw} by assuming that $M_R^\dagger M_R$ is non-singular
and that all its eigenvalues are of order $m_R^2$,
with $m_R \gg m_Z$. 
This leads to the effective Majorana mass matrix 
\begin{equation}\label{seesawMnu}
\mathcal{M}_\nu = - M_D^T M_R^{-1} M_D
\end{equation}
for the light neutrinos. In Eq.~(\ref{seesawMnu}),
$M_D$ is the Dirac mass matrix of the neutrinos. 
Allowing for an arbitrary number $n_H$ of Higgs doublets, we
avoid flavour-changing neutral Yukawa interactions by requiring
that all the Yukawa-coupling matrices be diagonal---hence,
$M_D$ too is diagonal.
This procedure is ``natural,''
since it amounts to conservation of the three lepton numbers 
$L_e$, $L_\mu$, and $L_\tau$ in the Lagrangian.
The only exception to this conservation is
the Majorana mass term in Eq.~(\ref{LM}),
where the lepton numbers are allowed to be broken \emph{softly}.
Despite the soft breaking of the lepton numbers
$L_\alpha$ ($\alpha = e, \mu, \tau$) at the high scale $m_R$,
the resulting theory is
well-behaved with respect to flavour-changing interactions and,
moreover, it exhibits an interesting non-decoupling of the neutral
scalar interactions for $m_R \to \infty$ when $n_H \geq 2$ \cite{GL02}.
In this framework,
maximal atmospheric neutrino mixing is implemented by the symmetry
\begin{equation}\label{Z2-MAMM}
\mathbbm{Z}_2: \quad
\nu_{\mu R} \leftrightarrow \nu_{\tau R}\, , \;
D_\mu \leftrightarrow D_\tau\, , \;
\mu_R \leftrightarrow \tau_R \,,
\end{equation}
where $D_\alpha$ denotes the left-handed lepton doublets.
Because of $\mathbbm{Z}_2$ we have
\begin{equation}
\left( M_R \right)_{e \mu} = \left( M_R \right)_{e \tau}\, , \quad
\left( M_R \right)_{\mu \mu} = \left( M_R \right)_{\tau \tau}\, ,
\label{jghsd}
\end{equation}
and
\begin{equation} \label{MD}
M_D = \mbox{diag} \left( a, b, b \right) \,.
\end{equation}
As a consequence,
the light-neutrino Majorana mass matrix of Eq.~(\ref{seesawMnu})
has the same structure as $M_R$:
\begin{equation}\label{Mnu}
\mathcal{M}_\nu =
\left( \begin{array}{ccc}
x & y & y \\ y & z & w \\ y & w & z
\end{array} \right).
\end{equation}
Maximal atmospheric neutrino mixing and $U_{e3} = 0$
immediately follow from this structure of $\mathcal{M}_\nu$. 
We stress that this structure results from a
symmetry and that the MAMM, therefore, is really a model in the
technical sense, not just a texture.\footnote{The charged-lepton mass matrix
remains diagonal because of the assumed conservation,
in all dimension-4 couplings,
of the three lepton numbers $L_\alpha$.}
Using an adequate phase convention and dropping possible Majorana phases,
from the $\mathcal{M}_\nu$ of Eq.~(\ref{Mnu}) we obtain the
lepton mixing matrix 
\begin{equation}\label{U}
U = \left( \begin{array}{ccc}
\cos \theta & \sin \theta & 0 \\
\sin \theta/\sqrt{2} & -\cos \theta/\sqrt{2} & -1/\sqrt{2} \\
\sin \theta/\sqrt{2} & -\cos \theta/\sqrt{2} &  1/\sqrt{2}
\end{array} \right).
\end{equation}
Since $m_\mu \neq m_\tau$, 
the $\mathbbm{Z}_2$ symmetry of Eq.~(\ref{Z2-MAMM})
must be broken spontaneously by the vacuum
expectation value (VEV) of some Higgs doublet transforming
non-trivially under $\mathbbm{Z}_2$. To avoid destruction of
the form in Eq.~(\ref{Mnu}) of the light-neutrino mass matrix, such a Higgs
doublet must not contribute to $M_D$, but only to the mass matrix of
the charged leptons. In Ref.~\cite{GL01} this problem was solved by 
having altogether three Higgs doublets and an additional
$\mathbbm{Z}'_2$ symmetry; since that solution cannot be directly
transferred to an $SU(5)$ model,
we shall not discuss it in detail here.

The $\mathbbm{Z}_2$ of Eq.~(\ref{Z2-MAMM}) does not commute with the
$U(1)$ associated with the lepton numbers $L_\mu$ and $L_\tau$. It is easy
to see that we have in the MAMM the horizontal non-abelian symmetry group
\begin{equation}
U(1)_{L_e} \times U(1)_{(L_\mu + L_\tau)/2} \times 
O(2)_{(L_\mu - L_\tau)/2} \,.
\end{equation}
We have indicated the lepton-number combinations associated with the
$U(1)$ groups;
the $O(2)$ is generated by
\begin{equation}
\left(\begin{array}{cc} 
e^{i\alpha (L_\mu - L_\tau)/2} & 0 \\
0 & e^{-i\alpha (L_\mu - L_\tau)/2}
\end{array} \right) \; (\alpha \in \mathbbm{R} ) 
\quad \mathrm{and} \quad
\left( 
\begin{array}{cc} 0 & 1 \\ 1 & 0 \end{array}
\right),
\end{equation}
corresponding to $U(1)_{(L_\mu - L_\tau)/2}$ and $\mathbbm{Z}_2$,
respectively. 

In this paper we shall discuss how the main features of the
MAMM, namely the groups $U(1)_{L_\alpha}$ softly broken by
$\mathcal{L}_M$ and the symmetry $\mathbbm{Z}_2$,
can be embedded in an $SU(5)$ GUT.
This will require a discussion of $\mathbbm{Z}_2$ breaking
within $SU(5)$
with the purpose of allowing for a non-trivial CKM matrix.
These subjects will be dealt with in
Section \ref{embedding}. Since we shall end up with a proliferation
of scalar multiplets,
and since our model is in principle non-supersymmetric,
gauge-coupling unification in the MAMM $SU(5)$ embedding
is a non-trivial undertaking. Possible solutions to this problem
will be studied in
Section~\ref{unification}. Finally, our conclusions are presented in
Section \ref{conclusions}. The two appendices contain $SU(5)$
technicalities; the Yukawa couplings of the scalar $SU(5)$-plets which
we need, and the charged-fermion mass matrices, are given
in Appendix~\ref{su5yukawa}; in Appendix~\ref{branch}
we collect the branching rules with respect to $G_\mathrm{SM}$ of
some irreducible representations (irreps) of $SU(5)$.

\section{Maximal atmospheric neutrino mixing in $SU(5)$}
\label{embedding}

\paragraph*{\boldmath $SU(5)$ \unboldmath\ preliminaries:}
The chiral fermion fields of one SM family
are accommodated in $SU(5)$ irreps in the following way \cite{su5,ross}: 
$\psi_R \sim \underline{5}$ and $\chi_L \sim \underline{10}$,
where the 10-plet is obtained as the antisymmetric part of 
$\underline{5} \otimes \underline{5}$. 
The $\underline{5}$,
which is the defining representation of $SU(5)$, 
has the generator of electric charge 
\begin{equation}
Q_{\underline{5}} = \mathrm{diag} \left( -1/3, -1/3, -1/3, +1, 0 \right).
\end{equation}
The fermion multiplets,
in terms of chiral SM fields,
are given by
\begin{equation}
\psi_R = \left( \begin{array}{c}
d^1_R \\ d^2_R \\ d^3_R \\ C \overline{\ell_L}^T \\
- C \overline{\nu_L}^T \end{array} \right),
\quad
\chi_L = \left( \begin{array}{ccccc}
0 & C \overline{u^3_R}^T & - C \overline{u^2_R}^T & -u^1_L & -d^1_L \\
-C \overline{u^3_R}^T & 0 & C \overline{u^1_R}^T & -u^2_L & -d^2_L \\
C \overline{u^2_R}^T & -C \overline{u^1_R}^T & 0 & -u^3_L & -d^3_L \\
u^1_L & u^2_L & u^3_L & 0 & -C \overline{\ell_R}^T \\
d^1_L & d^2_L & d^3_L & C \overline{\ell_R}^T & 0
\end{array} \right),
\label{chiL}
\end{equation}
where the upper indices $1,2,3$ are colour-$SU(3)$ indices.
Note that $\chi_L^{ij} = - \chi_L^{ji}$.

The scalar $SU(5)$ multiplets which may couple to fermionic bilinears
are determined by the following tensor products,
allowed by the chiral structure of $\psi_R$ and $\chi_L$ \cite{ross,slansky}:
\begin{eqnarray}
\underline{5} \otimes \underline{5} &=&
\underline{15} \oplus \underline{10}\, , \label{55} \\
\underline{5}^\ast \otimes \underline{10} &=& 
\underline{5} \oplus \underline{45}^\ast\, , \label{510} \\
\underline{10} \otimes \underline{10} &=&
\underline{5}^\ast \oplus \underline{50} \oplus \underline{45}\, .
\label{1010}
\end{eqnarray}
The only scalar multiplets needed for our Yukawa couplings
transform according to the irreps $\underline{5}$ and $\underline{45}$,
or their complex conjugates \cite{ross};
see Appendix~\ref{su5yukawa}
for the construction of their Yukawa-coupling Lagrangians.
In the following,
the scalar 45-plets will be distinguished from the scalar 5-plets by a tilde.

All fermionic multiplets appear threefold,
thus with family indices $a = 1,2,3$ they are denoted $\psi_{Ra}$,
$\chi_{La}$,
and $\nu_{Ra}$.
The right-handed neutrinos are $SU(5)$ singlets:
$\nu_{Ra} \sim \underline{1}$.

\paragraph*{The family number:}
Implementing the idea of Ref.~\cite{GL01},
we want to have $M_D$ (the neutrino Dirac mass matrix)
and $M_\ell$ (the charged-lepton mass matrix)
simultaneously diagonal.
This must be enforced by means of some symmetry.
If $M_\ell$ is diagonal because of a symmetry,
then we see from Eq.~(\ref{mell})
that the Yukawa-coupling matrices $Y_d$ and $\tilde Y_d$ must be diagonal;
but then,
from Eq.~(\ref{md}),
$M_d$ turns out diagonal too.
This means that quark mixing must stem exclusively from $M_u$,
in the same way that lepton mixing originates exclusively from $M_R$.

Let us assume that,
in analogy to the MAMM,
there is only one 5-plet $H_R$ coupling to the $\nu_R$.
Then we have the following terms in the Lagrangian
(see,
for instance,
Refs.~\cite{ross,buchmueller}):
\begin{equation}
\bar \nu_{Ra} C \left( \bar \psi_{Rbi} \right)^T \!
H_R^i \left( Y_\nu \right)_{ab}
- \frac{1}{2}\, \bar \nu_R M_R C {\bar \nu_R}^T + \mathrm{H.c.}
\label{nhurk}
\end{equation}
The seesaw mechanism is operative
and the light-neutrino mass matrix is given by Eq.~(\ref{seesawMnu}),
where
\begin{equation}\label{MD1}
M_D = v_R Y_\nu/\sqrt{2} \, ,
\end{equation}
where $v_R / \sqrt{2}$ is the VEV of $H_R$.
We introduce the family-number symmetry
\begin{equation}
F = \mathrm{diag} \left( 0, +1,-1 \right),
\end{equation}
applying both to the $\psi_{Ra}$ and to the $\chi_{La}$.
In the $\nu_{Ra}$ sector
one has $F = \mathrm{diag} \left( 0, -1, +1 \right)$ instead.
The scalar multiplet $H_R$
coupling to $\bar \nu_R C \bar \psi_R^T$
(see Eq.~(\ref{nhurk}))
and all the scalar multiplets coupling to $\bar \psi_R \chi_L$
(see Appendix~\ref{su5yukawa})
are assumed to have $F=0$.
The family numbers of the
scalar multiplets coupling to $\chi_L^T C^{-1} \chi_L$
(see Appendix~\ref{su5yukawa})
will be discussed later.
The symmetry group $U(1)_F$ defined here
overtakes the role of the three groups $U(1)_{L_\alpha}$ of the MAMM. 
As a consequence of the symmetry $U(1)_F$,
the Yukawa-coupling matrices $Y_\nu$, 
$Y_d$,
and $\tilde Y_d$ are all forced to be diagonal,
as we wanted;
$U(1)_F$ is softly broken by the Majorana mass terms
of the right-handed neutrinos,
i.e.\ by $M_R$.

\paragraph*{Maximal atmospheric neutrino mixing:}
In analogy to Eq.~(\ref{Z2-MAMM}),
we next introduce an interchange symmetry
between the second and third families:
\begin{equation}\label{Z2-MAM}
\mathbbm{Z}_2: \quad
\psi_{R2} \leftrightarrow \psi_{R3}\, , \quad
\chi_{L2} \leftrightarrow \chi_{L3}\, , \quad
\nu_{R2} \leftrightarrow \nu_{R3}\, .
\end{equation}
This forces $\left( Y_\nu \right)_{22} = \left( Y_\nu \right)_{33}$
and therefore leads to $M_D$ of the form in Eq.~(\ref{MD}).
The matrix $M_R$ moreover satisfies
$\left( M_R \right)_{12} = \left( M_R \right)_{13}$ and
$\left( M_R \right)_{22} = \left( M_R \right)_{33}$,
just as in Eq.~(\ref{jghsd}).
Therefore $\mathcal{M}_\nu$ is as in Eq.~(\ref{Mnu})
and we have maximal atmospheric neutrino mixing implemented.
Note that $U(1)_F$ together with the $\mathbbm{Z}_2$ of Eq.~(\ref{Z2-MAM})
generate a symmetry group $O(2)$.

\paragraph*{The down-type-quark and charged-lepton masses:}
We must check whether the introduction of $U(1)_F$
and of $\mathbbm{Z}_2$ is not incompatible with the freedom necessary
to accommodate all the charged-fermion masses and CKM mixing angles.
The CKM matrix is not the unit matrix,
therefore the up-type-quark mass matrix $M_u$ cannot be diagonal,
contrary to what happens with the down-type-quark mass matrix $M_d$;
this implies that we must allow for non-diagonal
Yukawa-coupling matrices for fermionic bilinears
of the type $\chi_L^T C^{-1} \chi_L$.
In order to obtain this,
it is useful to separate the scalar multiplets coupling to $\bar \psi_R \chi_L$
from those coupling to $\chi_L^T C^{-1} \chi_L$.
Furthermore,
as we shall see below,
in order to reproduce the down-type-quark masses
while avoiding destruction of the form of $M_D$ in Eq.~(\ref{MD}),
one also needs to ensure that $H_R$ is the only scalar multiplet
coupling to the $\nu_{Ra}$.
In order to reproduce the down-type-quark masses
and the charged-lepton masses
we need two 5-plets $H$ and $H'$ together with one 45-plet $\tilde H$.
We introduce the symmetries
\begin{eqnarray}
\mathbbm{Z}^\prime_2:  && \quad \nu_R \to -\nu_R \,, \quad H_R \to -H_R 
\label{Z'} \\
\mathbbm{Z}^{\prime \prime}_2: &&
\quad \chi_L \to -\chi_L\, ,
\quad H \to -H\, , \quad H^\prime \to - H^\prime\, , \quad
\tilde H \to - \tilde H\, ,
\label{Z''}
\end{eqnarray}
which allow couplings of $H_R$ only to $\bar \nu_R C \bar \psi_R^T$
(see Eq.~(\ref{nhurk})),
while $H$,
$H^\prime$,
and $\tilde H$ couple only to $\bar \psi_R \chi_L$. 
All the scalar multiplets coupling to $\chi_L^T C^{-1} \chi_L$,
and thereby generating $M_u$,
are invariant under both $\mathbbm{Z}^\prime_2$
and $\mathbbm{Z}^{\prime \prime}_2$.

We supplement the symmetry $\mathbbm{Z}_2$ of Eq.~(\ref{Z2-MAM}) with
\begin{equation}
\mathbbm{Z}_2: \quad H' \to -H' \,,
\end{equation}
while $H_R$,
$H$,
and $\tilde H$ transform trivially under $\mathbbm{Z}_2$.
Denoting the Yukawa-coupling matrices of $H$ and $H^\prime$
by $Y_d$ and $Y^\prime_d$,
respectively,
the symmetry $\mathbbm{Z}_2$ leads to
$\left( Y_d \right)_{22} = \left( Y_d \right)_{33}$,
$\left( Y_d^\prime \right)_{22}
= - \left( Y_d^\prime \right)_{33}$,
and $\left( Y_d^\prime \right)_{11} = 0$.
The Yukawa-coupling matrix $\tilde Y$ of $\tilde H$ satisfies
$\left( \tilde Y_d \right)_{22} = \left( \tilde Y_d \right)_{33}$.
Then,
with Eqs.~(\ref{md}) and (\ref{mell}) of Appendix~\ref{su5yukawa},
we obtain
\begin{eqnarray}
M_d &=& \mathrm{diag} \left( r+s, m+n+q, m-n+q \right), 
\label{Md} \\
M_\ell &=& \mathrm{diag} \left( r-3s, m+n-3q, m-n-3q \right),
\label{Mell}
\end{eqnarray}
where $r$, $s$, $m$, $n$, and $q$ are complex parameters.
This allows for the masses $m_d = |r+s|$ and $m_e = |r-3s|$
to be unrelated. 
As for $m_s$,
$m_b$,
$m_\mu$,
and $m_\tau$,
they are given by only three effective parameters:
$n$,
$m+q$,
and $m-3q$.
As $m_s \ll m_b$ and $m_\mu \ll m_\tau$,
we find $m+q \simeq -n \simeq m-3q$,
and this in turn leads to the approximate relation \cite{su5}
\begin{equation}\label{mtau=mb}
m_\tau/m_b \simeq 1\, ,
\end{equation}
which is valid at the GUT scale.
Clearly,
$m_s$ and $m_\mu$ remain unrelated.
It is well known that Eq.~(\ref{mtau=mb}) leads to the correct ratio
$m_\tau / m_b$ at low energies \cite{buras} (see also Ref.~\cite{inoue}).
Thus,
in our scheme one is able to accommodate all the down-type-quark
and charged-lepton masses,
and one is still rewarded with the correct relation~(\ref{mtau=mb})
at the GUT scale.

\paragraph*{The up-type-quark masses and the CKM angles:}
It remains to demonstrate that the known up-type-quark masses
and CKM matrix can be accommodated through $M_u$.
The scalar multiplets not coupling to $\psi_R$ and $\nu_R$
will be denoted $H_z$ (which are 5-plets)
and $\tilde H_z$ (45-plets).
They are invariant under both $\mathbbm{Z}^\prime_2$
and $\mathbbm{Z}^{\prime \prime}_2$.
First we consider the constraints from the family number $F$. 
The terms $\left( \chi^{ij}_{La} \right)^T \!
C^{-1} \chi^{kl}_{Lb}$ have $F$ quantum numbers given by the matrix
\begin{equation}\label{muF}
\left( \begin{array}{ccc}
0 & +1 & -1 \\ +1 & +2 & 0 \\ -1 & 0 & -2
\end{array} \right).
\end{equation}
Clearly,
in order for the CKM matrix to be non-trivial
we must allow for $H_z$ and $\tilde H_z$
to carry a non-zero family number $F = z$;
this means that the subscript $z$ gives,
by definition,
the $F$-value of the scalar multiplet.
For the 5-plets we have the possibilities $H_0$,
$H_{\pm 1}$,
and $H_{\pm 2}$;
whereas for the 45-plets $\tilde H_z$,
which couple through antisymmetric matrices,
only $z = 0$ and $z = \pm 1$ have an impact on $M_u$.
If,
for a given pair of family indices $(a,b)$
corresponding to a family number $F = -z$,
there is only $H_z$,
then we shall end up with $\left( M_u \right)_{ab} = \left( M_u \right)_{ba}$;
if,
on the contrary,
there is no $H_z$ but only $\tilde H_z$,
then we shall have $\left( M_u \right)_{ab} = - \left( M_u \right)_{ba}$;
if both $H_z$ and $\tilde H_z$ are present,
then the matrix elements $\left( M_u \right)_{ab}$
and $\left( M_u \right)_{ba}$ will be unrelated;
if neither $H_z$ nor $\tilde H_z$ exist,
then $\left( M_u \right)_{ab} = \left( M_u \right)_{ba} = 0$.

Let us now proceed to take into account the symmetry $\mathbbm{Z}_2$.
We consider the above scalar multiplets $H_{0, \pm 1, \pm 2}$
and $\tilde{H}_{0,\pm 1}$.
Under $\mathbbm{Z}_2$ we require that
\begin{equation}\label{HZ2}
\mathbbm{Z}_2 : \quad \left\{
\begin{array}{lll} 
H_0 \to H_0 \,, &
H_1 \leftrightarrow H_{-1} \,, &
H_2 \leftrightarrow H_{-2} \,, \\
\tilde{H}_0 \to -\tilde{H}_0 \,, &
\tilde{H}_1 \leftrightarrow \tilde{H}_{-1} \,. &
\end{array}
\right.
\end{equation}
We then find the following Yukawa couplings of the scalar 5-plets, 
compatible with $\mathbbm{Z}_2$ and with $F$:
\begin{eqnarray}
{\cal L}_Y^{5,u} &=& \epsilon_{ijklp} \left\{
\left( H_0 \right)^p
\left[ a \left( \chi_{L1}^{ij} \right)^T \! C^{-1} \chi_{L1}^{kl}
+ b \left( \chi_{L2}^{ij} \right)^T \! C^{-1} \chi_{L3}^{kl}
+ b \left( \chi_{L3}^{ij} \right)^T \! C^{-1} \chi_{L2}^{kl} \right]
\right. \nonumber \\ & &
+ c \left( \chi_{L1}^{ij} \right)^T \! C^{-1} \left[
\chi_{L2}^{kl} \left( H_{-1} \right)^p
+ \chi_{L3}^{kl} \left( H_1 \right)^p \right] 
\nonumber \\ & &
+ c  \left[
\left( \chi_{L2}^{ij} \right)^T \! \left( H_{-1} \right)^p
+ \left( \chi_{L3}^{ij} \right)^T \! \left( H_1 \right)^p \right] C^{-1}
\left( \chi_{L1}^{kl} \right)
\nonumber \\ & &
\left. + d \left[ 
\left( \chi_{L2}^{ij} \right)^T \! C^{-1} \chi_{L2}^{kl}
\left( H_{-2} \right)^p + 
\left( \chi_{L3}^{ij} \right)^T \! C^{-1} \chi_{L3}^{kl}
\left( H_2 \right)^p 
\right] \right\} + \mathrm{H.c.}
\label{Mu5}
\end{eqnarray}
With the 45-plets $\tilde{H}_{0, \pm 1}$
there are the following Yukawa couplings:
\begin{eqnarray}
\mathcal{L}_Y^{45,u} &=& \epsilon_{ijklp} \left\{
r \left( \tilde{H}_0 \right)^{lp}_q 
\left[
\left( \chi_{L2}^{ij} \right)^T \! C^{-1} \chi_{L3}^{kq}
- 
\left( \chi_{L3}^{ij} \right)^T \! C^{-1} \chi_{L2}^{kq} 
\right] \right. \nonumber \\ 
& & 
+ t \left( \chi_{L1}^{ij} \right)^T \! C^{-1} \left[ 
\chi_{L2}^{kq} \left( \tilde{H}_{-1} \right)^{lp}_q
+  \chi_{L3}^{kq} \left( \tilde{H}_1 \right)^{lp}_q \right]
\nonumber \\
& & \left.
- t \left[ 
\left( \chi_{L2}^{ij} \right)^T \! \left( \tilde{H}_{-1} \right)^{lp}_q
+ \left( \chi_{L3}^{ij} \right)^T \! \left( \tilde{H}_1 \right)^{lp}_q
\right] C^{-1} \chi_{L1}^{kq}  
\right\}
+ \mathrm{H.c.} \label{Mu45}
\end{eqnarray}
After the spontaneous breaking of $\mathbbm{Z}_2$,
the couplings in Eq.~(\ref{Mu5}) yield a symmetric $M_u$;
if we have both Eqs.~(\ref{Mu5}) and (\ref{Mu45}),
then we end up with a completely general up-type-quark mass matrix.

There is a lot of freedom in choosing among the possible scalar
multiplets which may contribute to $M_u$,
and one might think of deriving relations between the CKM mixing angles
and the up-type-quark mass ratios.
It is,
however,
difficult to imagine any such relation
which might turn out to be in agreement with the known values
of those quantities,
since the up-type-quark mass ratios are unfavourably small.
In the next section we shall simply assume a symmetric $M_u$
generated by the five scalar 5-plets $H_0$,
$H_{\pm 1}$,
and $H_{\pm 2}$,
thereby discarding any possible 45-plets.

The family number $F$ is softly broken,
through terms of dimension 3,
by the mass Lagrangian of the right-handed neutrino singlets;
consequently,
soft $F$-breaking terms must be considered also in the Higgs potential.
Below the $SU(5)$ scale,
$F$ is effectively conserved in the Yukawa couplings of the leptons;
at low scales its role is overtaken by the lepton numbers $L_\alpha$.
Thus, 
the idea of softly broken lepton numbers,
advocated in Ref.~\cite{GL01},
is compatible with an $SU(5)$ GUT.
The family number $F$ is also spontaneously broken, at the weak scale,
by the VEVs of the scalars with $F \neq 0$,
which are needed for reproducing the known up-type-quark masses
and CKM angles.

\section{Gauge-coupling unification}
\label{unification}

Since we are constructing an $SU(5)$ GUT,
we have to address the issue of gauge-coupling unification
and we must check that things can be arranged in such a way that
the unification scale $m_U$ lies in the range $10^{16}$ to $10^{19}$ GeV; 
the lower value is determined by the need to avoid proton decay,
the higher value corresponds to the Planck mass. We follow the strategy of
Refs.~\cite{giunti,fuerstenau,rizzo} and use the one-loop
renormalization-group equations (RGE) for the gauge couplings, 
which are decoupled from the RGE for the Yukawa couplings
and for the scalar-potential couplings---see, e.g., Ref.~\cite{cheng}.
We assume the ``desert'' hypothesis,
i.e.\ that there are no particles with masses in between the Fermi scale
(which we represent by the mass $m_Z$ of the Z boson)
and $m_U$.
We then have
\begin{eqnarray}
\frac{1}{\alpha_U}
&=& \omega_1 - \frac{t}{2 \pi}
\left( \frac{41}{10} + a_1 \right) \label{a1} \\
&=& \omega_2 - \frac{t}{2 \pi}
\left( - \frac{19}{6} + a_2 \right) \label{a2} \\
&=& \omega_3 - \frac{l}{2 \pi}
\left( - 7 + a_3 \right)\,. \label{a3}
\end{eqnarray}
In these equations,
$\alpha_U$ is the fine-structure constant
corresponding to the $SU(5)$ gauge coupling
at the scale $m_U$,
$t = \ln \left( m_U / m_Z \right)$,
and $\omega_j = 1 / \alpha_j \left( m_Z \right)$ for $j = 1, 2, 3$.
The numbers $41/10$,
$-19/6$,
and $-7$ in Eqs.~(\ref{a1})--(\ref{a3}) are the contributions to the RGE
from the SM multiplets \cite{cheng};
in particular,
the numbers $41/10$ in Eq.~(\ref{a1}) and $-19/6$ in Eq.~(\ref{a2})
include the effects of the single Higgs doublet of the SM.
The numbers $a_1$,
$a_2$,
and $a_3$ are the contributions to the RGE from any multiplets,
beyond the SM ones,
which might exist at (or below) the Fermi scale.
Using $\omega_1$ and $\omega_2$,
which are rather well known,
as inputs,
while $m_U$ and $\alpha_3(m_Z)$ are treated as outputs,
one derives from Eqs.~(\ref{a1})--(\ref{a3}) that
\begin{eqnarray}
\ln \left( m_U / m_Z \right) 
&=& \frac{30 \pi \left( \omega_1 - \omega_2 \right)}
{109 + 15 \left( a_1 - a_2 \right)}\, , \label{t} \\
\alpha_3(m_Z) &=& \frac{2 \left[ 109 + 15 \left( a_1 - a_2 \right) \right]}
{3 \omega_2 \left[ 111 + 10 \left( a_1 - a_3 \right) \right]
- 5 \omega_1 \left[ 23 + 6 \left( a_2 - a_3 \right) \right]}\, . \label{alfa3}
\end{eqnarray}

Numerically,
we use
\begin{equation}\label{input}
\alpha_3(m_Z) = 0.1200(28) \,, \quad
\hat \alpha(m_Z)^{-1} = 127.934(27) \,, \quad {\rm and}\ \quad
\sin^2 \hat \theta_w(m_Z) = 0.23113(15)\, ,
\end{equation}
from the article by Erler and Langacker in Ref.~\cite{PDG}.
In Eq.~(\ref{input}),
$\alpha$ is the fine-structure constant
and $\theta_w$ is the weak mixing angle; the hats
indicate that the $\overline{\mathrm{MS}}$ renormalization scheme
has been used in obtaining those quantities. 
Then, at the energy scale $m_U$, the values of 
\begin{equation}\label{alfa12}
\alpha_1 = \frac{5}{3}\, \frac{\alpha}{\cos^2 \theta_w} \,, \quad
\alpha_2 = \frac{\alpha}{\sin^2 \theta_w}\, ,
\end{equation}
and $\alpha_3$ become identical,
{\it cf.} Eqs.~(\ref{a1})--(\ref{a3}).
When applying Eqs.~(\ref{t}) and (\ref{alfa3}), 
we use as input the mean values of $\hat \alpha(m_Z)^{-1}$
and $\sin^2 \hat \theta_w \left( m_Z \right)$ in Eq.~(\ref{input}),
together with Eq.~(\ref{alfa12}),
for the computation of $\omega_1$ and $\omega_2$.

The scalar representations $\underline{5}$ and $\underline{45}^\ast$
of $SU(5)$ each contain one Higgs doublet $(1,2)_{1/2}$
(in the notation $(a,b)_c$ the numbers $a$ and $b$
are the dimensions of the representations of the $SU(3)$ and $SU(2)$
subgroups of $SU(5)$,
respectively,
while $c$ is the value of the weak hypercharge).
The VEVs of those Higgs doublets are of the order of the electroweak scale,
and therefore the masses of those Higgs doublets,
too,
are at the Fermi scale.
Since every $\underline{5}$ or $\underline{45}$ of $SU(5)$
supplies one light Higgs doublet, 
our model has (at least) nine Higgs doublets:
one each from $H_R$, $H$, $H^\prime$, $\tilde H$,
$H_0$, $H_{\pm 1}$, and $H_{\pm 2}$.
This makes eight low-mass Higgs doublets beyond the one in the SM.
From Table \ref{deltaa} we may compute
the corresponding contributions to the $a_j$;
one obtains $10 \left( a_1 - a_3 \right) = 6 \left( a_2 - a_3 \right)
= - 15 \left( a_1 - a_2 \right) = 8$.
Using Eqs.~(\ref{t}) and (\ref{alfa3}),
this leads to $m_U \simeq 8 \times 10^{13}$ GeV
and $\alpha_3 \left( m_Z \right) = 0.143$.
The latter value is not too far from what is required,
{\it cf.} Eq.~(\ref{input}),
but the GUT scale $m_U$ is much too low.
\begin{table}
\begin{center}
\begin{tabular}{l|ccc}
& $a_1$ & $a_2$ & $a_3$ \\
\hline
$(1,2)_{1/2}$ & 1/10 & 1/6 & 0 \\
$(6,2)_{-1/6}$ & 1/15 & 1 & 5/3 \\
$(1,2)_{3/2}$ & 9/10 & 1/6 & 0 \\
$(3,1)_{2/3}$ & 4/15 & 0 & 1/6
\end{tabular}
\end{center}
\caption{Contributions to $a_j$ ($j = 1,2,3$) of the $G_\mathrm{SM}$
  multiplets discussed in the text. \label{deltaa}}
\end{table}

As a consequence of the preceeding paragraph,
we need some additional multiplets of $G_{\rm SM}$ at the electroweak scale,
in particular some multiplets with non-trivial colour
which might shift $m_U$ to higher values
while keeping $\alpha_3 \left( m_Z \right)$ in the corrrect range.
Let us denote such a candidate $G_\mathrm{SM}$ multiplet by $D$
and investigate the conditions that we should impose on $D$.
To avoid problems with proton decay,
we require that $D$ be embedded in an $SU(5)$ irrep
which \emph{cannot} have any Yukawa couplings;
the lowest-dimensional eligible $SU(5)$ irreps 
are the $\underline{35}$ and the $\underline{40}$,
see Appendix~\ref{branch}.
Moreover,
since the scalars which have Yukawa couplings are 5 and 45-plets,
$D$ should not be contained in the decompositions of the $\underline{5}$,
the $\underline{45}$,
or their complex conjugates;
else,
$D$ might,
after the spontaneous breaking of the $SU(5)$ symmetry at $m_U$,
mix with analogous $G_\mathrm{SM}$ multiplets
from the $\underline{5}$ or $\underline{45}$,
and thereby end up having proton-decay-generating Yukawa couplings.
After imposing these two conditions,
we find that there are indeed some satisfactory candidates:
in particular,
the $(6,2)_{-1/6}$,
which is contained in both the $\underline{35}$
and the $\underline{40}$ of $SU(5)$,
and the $(1,2)_{3/2}$,
which is contained in the $\underline{40}$ (see Appendix~\ref{branch}).
The contributions of these multiplets of $G_\mathrm{SM}$ to the
$a_j$ are given in Table~\ref{deltaa}.
In particular,
we find that if,
beyond the nine Higgs doublets,
there are at the electroweak scale two $(6,2)_{-1/6}$ and one $(1,2)_{3/2}$,
then
\begin{equation}\label{MSSM}
a_1 - a_2 = - 5/3\, , \quad
a_1 - a_3 = - 3/2\, , \quad
a_2 - a_3 = 1/6\, .
\end{equation}
Using Eqs.~(\ref{t}) and (\ref{alfa3}),
we obtain the numerical result
\begin{equation}\label{res1}
m_U = 2 \times 10^{16} \; \mathrm{GeV}, \quad
\alpha_3 \left( m_Z \right) = 0.117\, .
\end{equation}
This demonstrates that we may achieve a sufficiently high GUT scale and,
simultaneously,
reproduce $\alpha_3 \left( m_Z \right)$ rather well.
We stress that the $(6,2)_{-1/6}$ and $(1,2)_{3/2}$
do not occur in the decomposition of any of the $SU(5)$ irreps
possibly coupling to fermions---see Eqs.~(\ref{55})--(\ref{1010})
and Appendix~\ref{branch}.
Therefore,
couplings of the $(6,2)_{-1/6}$ and the $(1,2)_{3/2}$ to the SM fermions
can only be induced by loop effects after $G_\mathrm{SM}$ breaking,
and it is justified to assume that any such couplings will be very small.
We need two light $(6,2)_{-1/6}$ and one light $(1,2)_{3/2}$,
which we may take, for instance,
from one $\underline{35}$ together with one $\underline{40}$ of $SU(5)$.
The other $G_\mathrm{SM}$ multiplets
in the $\underline{35}$ and $\underline{40}$ will have to be heavy,
with masses of order $m_U$.
This certainly means a fine-tuning problem for our theory,
analogous to the well-known doublet--triplet splitting
of the scalar 5-plets. 

It is well known that gauge-coupling unification in the MSSM
is compatible with the input data in Eq.~(\ref{input})
\cite{giunti,fuerstenau}.  
The numbers in Eq.~(\ref{MSSM}),
which determine $m_U$ and $\alpha_3 \left( m_Z \right)$,
are remarkable because they are exactly the same as in 
the Minimal Supersymmetric Standard Model (MSSM)
\cite{SUSY-SU5}.\footnote{The individual $a_j$ are different, however.}
The SM with eight extra Higgs doublets,
two $(6, 2)_{-1/6}$,
and one $(1,2)_{3/2}$
produces exactly the same $m_U$ and $\alpha_3 \left( m_Z \right)$
as the MSSM,
if we confine ourselves to the one-loop RGE.
Differences will arise only at the two-loop level.
Thus,
the gauge-coupling unification of the MSSM
can be imitated by simply adding a few scalar multiplets to the SM.

The choice of the $G_\mathrm{SM}$ multiplets
$(6,2)_{-1/6}$ and $(1,2)_{3/2}$
displays one additional noteworthy feature.
Let us assume that the total number of Higgs doublets is nine
(including the SM doublet)
and that the number of multiplets $(1,2)_{3/2}$ is one,
but let us allow the number $n_6$ of multiplets $(6,2)_{-1/6}$ to vary.
It turns out that in this case $\alpha_3 \left( m_Z \right)$
is independent of $n_6$ and is always given by
\begin{equation}\label{al3}
\alpha_3 \left( m_Z \right) = \frac{7}{12 \omega_2 - 5 \omega_1} \,.
\end{equation}
Thus,
numerically,
the value given in Eq.~(\ref{res1}) is precisely 
obtained from this formula,
which is also valid for the one-loop RGE result of the MSSM.
On the other hand, 
$m_U$ does depend on $n_6$,
which can be chosen such that $m_U$ lies in the correct range
without putting at peril the good Eq.~(\ref{al3}).
The best choice is $n_6 = 2$,
with the value of $m_U$ given in Eq.~(\ref{res1}). 
For $n_6 = 1$ the GUT scale comes down to $1.6 \times 10^{15}$ GeV,
which would result in much too fast proton decay,
whereas for $n_6 = 3$ it increases to $1.2 \times 10^{21}$ GeV,
above the Planck mass.

We may replace the $(1,2)_{3/2}$ by a $(3,1)_{2/3}$;
for the contributions to the $a_j$ see again Table~\ref{deltaa}. 
Then,
analogous to Eq.~(\ref{res1}),
we obtain
\begin{equation}\label{res2}
m_U = 4.1 \times 10^{17} \; \mathrm{GeV}, \quad
\alpha_3 \left( m_Z \right) = 0.117\, .
\end{equation}
$m_U$ is now higher than before,
while $\alpha_3 \left( m_Z \right)$ remains the same---by sheer coincidence,
its value is again given by Eq.~(\ref{al3}).
We note that with two $(6,2)_{-1/6}$
and two $(3,1)_{2/3}$ instead of one,
we can even allow for eleven light Higgs doublets,
with the result  $m_U = 1.7 \times 10^{17}$ GeV
and $\alpha_3 \left( m_Z \right) = 0.123$.
Using the $(3,1)_{2/3}$ instead of the $(1,2)_{3/2}$ makes a difference,
though.
The $(3,1)_{2/3}$ does not only occur
in the $G_{\rm SM}$ decomposition of the $\underline{40}$,
it also occurs in the $\underline{10}^\ast$. 
This means that,
after $SU(5)$ breaking,
a coupling of the light scalar multiplet $(3,1)_{2/3}$
to the fermionic bilinear $d^T_R C^{-1} d_R$ becomes allowed.
Such a coupling will be induced at loop level.
However,
it depends on the coupling strength of the $\underline{40}$
to the 5 and 45-plets in the Higgs potential
and it may in principle be made sufficiently small.

As for our simple usage of the one-loop RGE,
refinements are, of course, possible.
In particular,
we might use the two-loop RGE,
thereby taking into account the effect
of the large Yukawa coupling of the top quark.
It would also be possible to allow the light scalar multiplets
to be somewhat heavier than $m_Z$,
for instance with masses of order $0.5$ or $1$ TeV,
like in the MSSM.
Still another possibility would be to take into account
various threshold effects at the scale $m_U$.
Still,
the short study above shows that there certainly are acceptable ways
of making the gauge coupling constants unify in our $SU(5)$ theory
with nine light Higgs doublets.

\section{Conclusions} \label{conclusions}

In Ref.~\cite{GL01} a simple extension of the lepton sector of the SM
was put forward,
with three right-handed neutrino singlets and the seesaw mechanism,
and three Higgs doublets instead of one.
By requiring conservation of the three lepton numbers in the Yukawa sector,
while allowing them to be broken softly
by the Majorana mass terms of the right-handed singlets,
it was possible to enforce maximal atmospheric neutrino mixing
by means of a $\mathbbm{Z}_2$ symmetry,
while having arbitrary but in general large solar neutrino mixing.
Since in this model maximal atmospheric neutrino mixing
is enforced by means of a symmetry,
the value $45^\circ$ for the mixing angle
is stable under radiative corrections. 

In the present paper we have shown
that the suggestion of Ref.~\cite{GL01} can be embedded in $SU(5)$ GUTs.
Here we summarize the main features of the embedding:
\begin{itemize}
\item
Lepton mixing stems exclusively from the mass matrix $M_R$
of the right-handed singlets $\nu_R$;
atmospheric mixing is maximal;
the solar mixing angle is free in general---without fine-tuning
it will be large but not maximal;
$U_{e3} = 0$.
These are precisely the features of the tree-level mass matrix
found in the model of Ref.~\cite{GL01} which,
as we have now demonstrated,
can be transferred to $SU(5)$ GUTs.
\item
The CKM matrix is generated in the up-type-quark sector,
while the down-type-quark mass matrix is diagonal.
This is a consequence of the multiplet structure of $SU(5)$,
in particular,
of the 5-plet $\psi_R$ in Eq.~(\ref{chiL}).
\item
The family-number symmetry $U(1)_F$,
which is responsible for the diagonal character
of the matrices $M_D$ in Eq.~(\ref{MD1}),
$M_d$ in Eq.~(\ref{Md}),
and $M_\ell$ in Eq.~(\ref{Mell}),
is broken in two ways:
\emph{soft} breaking by $M_R$
and by terms of dimension two and three in the Higgs potential,
and \emph{spontaneous} breaking by the VEVs of the scalar multiplets
responsible for the up-type-quark mass matrix.
The non-trivial CKM matrix
is obtained via the spontaneous breaking of $U(1)_F$,
and the non-trivial lepton-mixing matrix
is obtained via the soft breaking of $U(1)_F$.
\item
On the other hand,
$\mathbbm{Z}_2$,
which is responsible for maximal atmospheric neutrino mixing
once the charged-lepton mass matrix is diagonal,
is broken only by the VEVs.
\item
In the lepton sector, the 
$U(1)_F$ enforces diagonal Yukawa couplings. Therefore, 
in this sector and below the GUT scale, instead of the family number $F$ we
have the usual three lepton numbers $L_\alpha$ ($\alpha = e, \mu, \tau$), 
which are only \emph{softly} broken by
the mass terms of the right-handed singlets \cite{GL01,GL02}. 
\end{itemize}
These features are probably the most generic ones of our model. 
The extension of the model of Ref.~\cite{GL01} to an $SU(5)$ GUT
is certainly not unique,
and the discussion in this paper should be perceived
as just an existence proof for that extension.
Among the features which might depend on the way
the extension is performed,
we may count the following ones:
\begin{itemize}
\item
The assignment of family numbers chosen by us,
together with our $\mathbbm{Z}_2$ symmetry,
can be thought of as originating from a non-abelian $O(2)$ symmetry group.
\item
Our choice of scalars coupling to the down-type-quark sector
produces the successful relation $m_b \simeq m_\tau$ at the GUT scale.
\item
We need eight scalar 5-plets and one 45-plet
in order to accommodate maximal atmospheric neutrino mixing,
the charged-fermion masses,
and CKM mixing.
It is,
therefore,
natural to assume that in our model there are nine light Higgs doublets.
We have shown that it is nevertheless possible
to obtain gauge-coupling unification,
even when assuming a desert between the electroweak and GUT scales.
\end{itemize}

We want to stress that obtaining gauge-coupling unification
in our model is not trivial at all.
However,
we were able to find an excellent solution,
in which the light scalar multiplets are nine Higgs doublets $(1,2)_{1/2}$,
one doublet $(1,2)_{3/2}$,
and two $(6,2)_{-1/6}$.
It is most remarkable that in this case the one-loop RGE
for the gauge couplings lead to results identical to those in the MSSM. 

Finally,
we remark that in our embedding
the up-type-quark mass matrix is a general symmetric mass matrix,
with no relationships among the up-type-quark masses and the CKM angles;
it is possible that more predicitive embeddings exist.

\section*{Acknowledgement}
\noindent
W.G.\ thanks H.\ Stremnitzer for many helpful discussions on $SU(5)$.

\newpage

\begin{appendix}
\setcounter{equation}{0}
\renewcommand{\theequation}{\Alph{section}\arabic{equation}}

\section{Yukawa couplings}
\label{su5yukawa}

In the following,
indices which transform through the
matrix $U \in SU(5)$ are written as upper indices;
indices transforming through the matrix $U^\ast$
are written as lower indices \cite{ross}.
It is clear from Eqs.~(\ref{55}),
(\ref{510}),
and (\ref{1010})
that the scalars which may have Yukawa couplings to $\psi_R$ and/or $\chi_L$
must be in one of the following representations of $SU(5)$:
$\underline{5}$,
$\underline{10}$,
$\underline{15}$,
$\underline{45}$,
or $\underline{50}$. 
We want to avoid spontaneous violation of colour or electric charge,
and to disallow a Majorana mass term for the left-handed neutrinos.
The scalar multiplets present in Yukawa couplings are,
therefore \cite{ross},
$H^i \sim \underline{5}$ and $\tilde{H}^{ij}_k \sim \underline{45}^\ast$;
the latter satisfies $\tilde{H}^{ij}_k= - \tilde{H}^{ji}_k$
and $\tilde{H}^{ij}_i = 0$
(we use the summation convention).
The Yukawa couplings are given by \cite{ross}
\begin{equation}
\mathcal{L}_\mathrm{Yukawa} = \mathcal{L}_Y^5 + \mathcal{L}_Y^{45}\, ,
\end{equation}
with
\begin{eqnarray}
\mathcal{L}_Y^5 &=&
\bar \psi_{Ria}\, \chi^{ij}_{Lb}\, H^\ast_j \left( Y_d \right)_{ab} -
\frac{1}{8}\, \epsilon_{ijklp} \left( \chi^{ij}_{La} \right)^T \! C^{-1}
\chi^{kl}_{Lb}\, H^p \left( Y_u \right)_{ab} + \mathrm{H.c.} \,, \label{L5} \\
\mathcal{L}_Y^{45} &=&
\frac{1}{2}\, \bar \psi_{Ria}\, \chi^{jk}_{Lb}\, \tilde{H}^{\ast i}_{jk}
\left( \tilde Y_d \right)_{ab} - \frac{1}{8}\, 
\epsilon_{ijklp} \left( \chi^{ij}_{La} \right)^T \! C^{-1} \chi^{kq}_{Lb}\,
\tilde{H}^{lp}_q \left( \tilde Y_u \right)_{ab} + \mathrm{H.c.}
\label{L45}
\end{eqnarray}
The numerical factors in these equations are conventional.
The symbol $\epsilon_{ijklp}$ represents the completely antisymmetric tensor,
which is normalized through $\epsilon_{12345} = +1$.
The indices $a$ and $b$ are flavour indices.
The Yukawa-coupling matrices $Y_d$ and $\tilde Y_d$
are general complex $3 \times 3$ matrices;
the matrix $Y_u$ is symmetric without loss of generality,
while $\tilde Y_u$ is antisymmetric:
\begin{equation}
\left( Y_u \right)_{ab} = \left( Y_u \right)_{ba}
\quad \mathrm{and} \quad
\left( \tilde Y_u \right)_{ab} = - \left( \tilde Y_u \right)_{ba}\, .
\end{equation}

The vacuum expectation values are given by \cite{ross}
\begin{equation}\label{VEV5}
\left\langle H^i \right\rangle_0 = \frac{v}{\sqrt{2}}\, \delta^i_5
\end{equation}
and
\begin{equation}\label{VEV45}
\left\langle \tilde{H}^{i5}_k \right\rangle_0 =
- \left\langle \tilde{H}^{5i}_k \right\rangle_0 =
\frac{\tilde v}{\sqrt{2}} 
\left( \delta^i_k - 4 \delta^i_4 \delta^4_k \right)
\; \mathrm{for} \; i \leq 4\,, \quad
\left\langle \tilde H^{ij}_k \right\rangle_0 = 0
\; \mathrm{for} \; i,j \leq 4\,.
\end{equation}

The charged-fermion mass matrices are defined by
\begin{equation}
\mathcal{L}_{\rm mass} = - \bar u_R M_u u_L - \bar d_R M_d d_L
- \bar \ell_R M_\ell \ell_L + {\rm H.c.}
\end{equation}
The relations
\begin{eqnarray}
\epsilon_{ijk45} \left( \chi^{ij}_{La} \right)^T \!
C^{-1} \chi^{k4}_{Lb} &=&
2\, \bar u_{Ra} u_{Lb}\, , \\
\label{auxrel}
\epsilon_{ijkl5} \left( \chi^{ij}_{La} \right)^T \!
C^{-1} \chi^{kl}_{Lb} &=&
4 \left( \bar u_{Ra} u_{Lb} + \bar u_{Rb} u_{La} \right),
\end{eqnarray}
which follow directly from the components of $\chi_L$
given in Eq.~(\ref{chiL}),
are useful for the extraction of the up-type-quark mass matrix $M_u$.
One obtains
\begin{eqnarray}
M_u &=&
\frac{1}{\sqrt{2}} \left( v Y_u - 2 \tilde v \tilde Y_u \right),
\label{mu} \\
M_d &=&
\frac{1}{\sqrt{2}} \left( v^\ast Y_d + \tilde{v}^\ast \tilde Y_d \right),
\label{md} \\
M_\ell &=&
\frac{1}{\sqrt{2}} \left( v^\ast Y_d^T - 3 \tilde{v}^\ast \tilde Y_d^T \right).
\label{mell}
\end{eqnarray}

\section{Branching rules}
\label{branch}

In this appendix we display the branching rules
for some representations of $SU(5)$
in terms of representations of $G_\mathrm{SM}$.
For simplicity we do not underline
the dimensions of the representations of $SU(3)$ and $SU(2)$.
The weak hypercharge $Y$ is normalized in the usual SM way,
i.e.\ $Y = Q - T_3$.
The branching rules below may,
for instance,
be found in Ref.~\cite{slansky}.\footnote{Note,
however,
that the irreps $\underline{35}$,
$\underline{40}$,
$\underline{45}$,
and $\underline{50}$ used here 
correspond to their respective complex conjugates in Ref.~\cite{slansky};
in the definitions of $\underline{45}$ and $\underline{50}$
we have followed Ref.~\cite{ross}.}

The defining representation of $SU(5)$ is
\begin{equation}\label{5branch}
\underline{5} = \left( 3, 1 \right)_{-1/3}
+ \left( 1, 2 \right)_{1/2}\, . 
\end{equation}
The product of two $\underline{5}$'s yields
\begin{eqnarray}
\underline{15} &=& \left( 6, 1 \right)_{-2/3} + \left( 3, 2 \right)_{1/6}
+ \left( 1, 3 \right)_{1}\, , \label{15branch} \\
\underline{10} &=& \left( 3^\ast, 1 \right)_{-2/3} + \left( 3, 2 \right)_{1/6}
+ \left( 1, 1 \right)_{1}\, . \label{10branch}
\end{eqnarray}
The representations $\underline{45}$ and $\underline{50}$ of $SU(5)$,
which arise in the tensor products of fermionic representations
in Eqs.~(\ref{510}) and (\ref{1010}),
have the following branching rules:
\begin{eqnarray}
\underline{45} &=& \left( 3, 1 \right)_{-4/3}
+ \left( 8, 2 \right)_{-1/2}
+ \left( 1, 2 \right)_{-1/2}
+ \left( 6, 1 \right)_{1/3}
+ \left( 3^\ast, 3 \right)_{1/3}
+ \left( 3^\ast, 1 \right)_{1/3}
\nonumber \\ & & + \left( 3, 2 \right)_{7/6}\, , \label{45branch} \\
\underline{50} &=& \left( 6^\ast, 1 \right)_{-4/3}
+ \left( 8, 2 \right)_{-1/2}
+ \left( 6, 3 \right)_{1/3}
+ \left( 3^\ast, 1 \right)_{1/3}
+ \left( 3, 2 \right)_{7/6}
+ \left( 1, 1 \right)_{2}\, . \label{50branch}
\end{eqnarray}
Finally, in Section~\ref{unification} we use the irreps
\underline{35} and \underline{40} and their decompositions:
\begin{eqnarray}
\underline{35} &=& \left( 10, 1 \right)_{-1} + \left( 6, 2 \right)_{-1/6}
+ \left( 3, 3 \right)_{2/3} + \left( 1, 4 \right)_{3/2}\, , \label{35} \\
\underline{40} &=& \left( 8, 1 \right)_{-1} + \left( 6, 2 \right)_{-1/6}
+ \left( 3^\ast, 2 \right)_{-1/6}
+ \left( 3, 3 \right)_{2/3} + \left( 3, 1 \right)_{2/3}
+ \left( 1, 2 \right)_{3/2}\, . \label{40}
\end{eqnarray}

\end{appendix}

\end{document}